\documentclass[final,3p,times,twocolumn]{elsarticle}

\usepackage{amssymb}
\usepackage{amsfonts}
\usepackage{graphicx}
\graphicspath{{Figures/}}
\usepackage{color}
\usepackage[normalem]{ulem}

\journal{Physics Letters A}

\begin{document}

\begin{frontmatter}



\title{On Improving the Performance of Nonphotochemical Quenching in 
CP29 Light-Harvesting Antenna Complex}


\author[label1]{Gennady P.  Berman} 

\address[label1]{Theoretical Division, T-4, Los Alamos National Laboratory, 
and the New Mexico Consortium,  Los Alamos, NM 87544, USA}

\author[label2]{Alexander I. Nesterov} 

\address[label2]{Departamento de F{\'\i}sica, CUCEI, Universidad de 
Guadalajara, Av. Revoluci\'on 1500, Guadalajara, CP 44420, Jalisco, M\'exico}

\author[label3]{ Richard T. Sayre}

\address[label3]{Biological Division, B-11, Los Alamos National Laboratory, 
and the New Mexico Consortium, Los Alamos, NM 87544, USA}

\author[label4]{Susanne Still}

\address[label4]{Department of Information and Computer Sciences, and 
Department of Physics and Astronomy, University of 
Hawaii at M\=anoa, 1860 East-West Road, Honolulu, HI 96822, USA}

\begin{abstract}
We model and simulate the performance of charge-transfer in nonphotochemical quenching (NPQ) in the CP29 light-harvesting antenna-complex associated with photosystem II (PSII). 
The model consists of five discrete excitonic energy states and two sinks, responsible for the potentially damaging processes and charge-transfer channels, respectively. We demonstrate that by varying (i) the parameters of the chlorophyll-based dimer, (ii) the resonant properties of the protein-solvent environment interaction, and (iii) the energy transfer rates to the sinks, one can significantly improve the performance of the NPQ. Our analysis suggests strategies for improving the performance of the NPQ in response to environmental changes, and may stimulate experimental verification.
\end{abstract}


\end{frontmatter}

\section{Introduction}
Photosynthesis in plants and algae is powered by rapid transfer of excitation energy to the reaction center (RC). At full sunlight intensities the rate of photon capture exceeds the rate of downstream electron transfer catalyzed by the cytochrome b6f complex by a factor of 10 \cite{Perrine}. 
 In this case, the PSII RC becomes over-reduced blocking further photochemistry. As a result, damaging processes occur. In particular, the energy of excess chlorophyll excited states can be used for production of singlet oxygen species which can destroy the photosynthetic organism.
To survive intense sunlight fluctuations, photosynthetic organisms have evolved many protective strategies, including NPQ \cite{book1} (and references therein).

NPQ is a strategy for partial suppression of the damaging channels. Excessive sunlight energy is transferred into quenching channels, including energy dissipation by the xanthophyll cycle carotenoids. The initiation of NPQ includes four main stages: (1) protonation processes (on timescales up to a few milliseconds) which are accompanied by decreasing pH in particular regions of the LHC; (2) geometrical reorganizations due to conformational changes of the protein-solvent environment (on timescales up to a few minutes and more); and (3) very rapid exciton transfer (ET) 
 to charge transfer or damaging channels, on timescales up to a few picoseconds.

Here, we consider only stage (3) of the NPQ process: energy transfer inside the LHC and to the sinks. The success of NPQ depends on the efficiency with which energy gets transferred away from damaging processes. In this paper, we ask how this efficiency depends, on one hand, on parameters that can be changed by the plant via adaptation (mutation), and, on the other hand, on noise characteristics of the protein-solvent environment. Specifically, we analyze stage (3) of the NPQ mechanism in CP29 LHCs of plants and green algae by a charge-transfer state (CTS), discussed in \cite{Ahn} (see also \cite{Dreuw1,Dreuw2,Ahn2})\footnote{Note, that there are many NPQ related mechanisms discussed in the literature, but a consensus does not yet exist, see e.g., \cite{book1}, and references therein.}. We build on the model developed in \cite{BNGS,BNLS}, and demonstrate numerically that the performance of the NPQ mechanism can be significantly improved by varying (i) characteristics of the chlorophyll dimer, (ii) resonant properties of protein-solvent interaction, and (iii) the ET rates to sinks. 
Our analysis suggests strategies for restoring NPQ efficiency in response to environmental changes.

\section{Model and main equations}
Energy transfer processes of the NPQ mechanism can be modeled by describing the sites associated with light-sensitive chlorophyll or carotenoid molecules by discrete excitonic energy states, $|n\rangle$ (where $n$ enumerates the sites of the LHC). Both the damaging and the quenching channels can be characterized by their corresponding energy sinks, $|S_n\rangle$, that provide independent continuum electron energy spectra \cite{BNGS,BNLS}. These sinks can have very complex structures. They can be responsible for many quasi-reversible chemical reactions, such as primary charge separation processes in the photosynthetic RCs \cite{Pudlak1,Pudlak2}, creation of CTS and singlet oxygen production \cite{BNGS,BNLS}, and coherent quantum effects \cite{Ber3,Ber4,Lloyd1,CDCH} even at ambient temperature. Generally, each sink, $|S_n\rangle$, is connected to a particular site, $|n\rangle$. The energy transfer from this state to its sink is characterized by the corresponding ET rate,  $\Gamma_n$. For those sites which are not connected to sinks, the corresponding ET rates vanish. 

\begin{figure}[ht]
 	\begin{center}
 		\scalebox{0.275}{\includegraphics{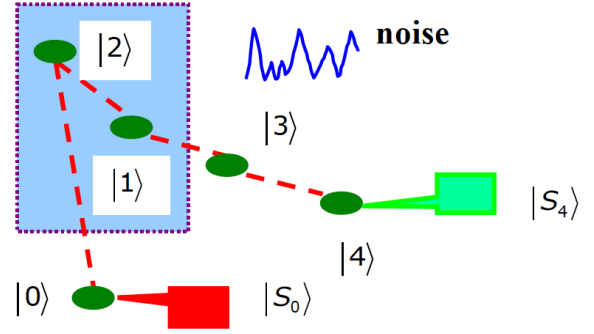}}
 	\end{center}
 	\caption{Schematic of the NPQ model consisting of five discrete excitonic 
 	states, $|n\rangle$, $(n=0,...,4)$, and two independent sinks, $|S_0 
 	\rangle$ (connected to the damaging state), and  $|S_4 \rangle$ 
 	(connected to the CTS that helps dissipate energy as part of the NPQ 
 	strategy). The blue rectangular indicates the dimer based on the excited 
 	states of a pair of chlorophyll molecules. The red dashed lines indicate 
 	non-zero matrix elements used in numerical simulations. 
 		\label{M}}
 \end{figure}

In our model, five discrete excitonic energy states and two energy sinks are embedded in a protein-solvent environment. The environment is modeled by a random telegraph process (RTP). (See Fig. \ref{M}.) For simplicity, we include only the first excited states of the chlorophyll and carotenoid molecules. More general consideration should also include their long-lived excited states \cite{book1}. This generalization is straightforward within our framework.
 
Under reasonable assumptions, the quantum dynamics of the ET can be described by an effective non-Hermitian Hamiltonian \cite{BNGS,BNLS},
   $ \tilde{\mathcal H}= {\mathcal H}- i \mathcal W$, where,
   \begin{eqnarray}\label{Ham}
   {\mathcal H} = \sum_{n}\varepsilon_n |n\rangle \langle n|  
   +\sum_{m\neq n} V_{mn} |m\rangle \langle n|, \nonumber \\
    m,n =0,1,\dots, 4,
   \end{eqnarray}
  is the dressed Hamiltonian, and
   \begin{eqnarray}
   \label{Gamma}
  {\mathcal W} = \frac{1}{2}\sum_{n=\{0,4\}}\Gamma_n|n\rangle \langle n|.
   \end{eqnarray}
 In (\ref{Ham}), $\varepsilon_n$ is the renormalized energy of the discrete state,  $|n\rangle$,	
   	and the parameter, $\Gamma_n$, in (\ref{Gamma})  is the tunneling rate 
   	to the $n$-{th} sink (only $\Gamma_0$ and $\Gamma_4$ are 
   	included).

  In our model, the dimer, which consists of two chlorophylls, $Chla_5$ and $Chlb_5$, each in an excited electron state, participates in the NPQ process, as described in \cite{Ahn}. 
  We use the following notation (see Fig. \ref{M} and \cite{Ahn}): The discrete  electron state, 
  $|1\rangle\equiv |Chla_5^*\rangle$, is the excited electron state of 
  $Chla_5$.  The discrete  electron state, $|2\rangle\equiv 
  |Chlb_5^*\rangle$, is the excited electron state of $Chlb_5$.  
  The discrete state,  $|3\rangle\equiv |(Chla_5-Zea)^*\rangle$, is the 
  heterodimer excited state, where ``Zea" denotes the carotenoid Zeaxanthin.  The discrete state,  $|4\rangle\equiv 
  |(Chla_5^{-}-Zea^{+})^*\rangle$, is the CTS of the heterodimer. The sink, $|S_4\rangle$, is the continuum part of the CTS 
  (channel), associated with dissipation of energy, and thereby suppression of the damaging 
  channel. The discrete part of the damaging  
  channel is the state, $|0\rangle$, and the sink, $|S_0\rangle$, is the continuum part of the damaging 
  channel.
  
   All matrix elements, $V_{mn}$, in the Hamiltonian 
  (\ref{Ham}) describe interactions between the discrete excitonic states. 
  Each sink is characterized by two parameters: the rate, $\Gamma_{0}$ 
  ($\Gamma_{4}$), of  ET into the sink, $|S_{0}\rangle$ ($|S_{4}\rangle$), 
  from the corresponding attached discrete state, $|0\rangle$ ($|4\rangle$), 
  and the efficiency (cumulative time-dependent probability), $\eta_{0}(t)$ 
  ($\eta_{4}(t)$), for the exciton to be absorbed by the corresponding sink.  
  Note, that both, $\Gamma_0$ and $\Gamma_4$, characterize only the rates 
  of destruction of the exciton in the CP29 LHC. They do not describe any 
  subsequent chemical reactions that take place in the damaging and the 
  charge-transfer channels. In this sense, our model describes only the 
  primary NPQ processes in the ET, and it does not describe the processes 
  which occur in both sinks. The latter occur on relatively large timescales, 
  and require a detailed knowledge of the structures of the sinks, and 
  additional methods for their analysis.

The dynamics of the system can be described by the Liouville-von Neumann equation,  \begin{eqnarray}\label{DM1}
    \dot{ \rho} = i[\rho,\mathcal H] - \{\mathcal W,\rho\},
 \end{eqnarray}
 where $\{\mathcal W,\rho\}= \mathcal W\rho +\rho\mathcal  W$.

We define the ET efficiency of tunneling to all $N$ sinks as,
\begin{eqnarray}\label{ET1} 
\eta(t) = 1 - {\rm Tr}(\rho(t)) =  \int_0^t {\rm Tr}\{\mathcal W,\rho(\tau)\} d \tau.
\end{eqnarray}
This can be expressed as the sum of time-integrated probabilities of trapping an electron into the $n$-th sink, \cite{Lloyd1,CDCH}:
\begin{eqnarray}
\eta(t) &=& \sum_{n}\eta_n(t), \\
\eta_n(t) &=& \Gamma_n \int_0^t \rho_{nn}(\tau)d \tau~,
\label{Eq16ar}
\end{eqnarray}
where $\eta_n(t)$ is the efficiency of the $n$-th sink. In particular, for our model (Fig. \ref{M}), complete suppression of the damaging channel {\it at all times} occurs if  $\eta_0(t)=0$. This would be the most desirable outcome, leading to maximal effectiveness of the NPQ mechanism.
 
 In the presence of the protein-solvent noisy environment, the evolution of the system can be described by the following effective non-Hermitian Hamiltonian \cite{BNGS,BNLS}, 
 \begin{eqnarray}
 \tilde{\mathcal H}_{tot}= {\mathcal H}- i \mathcal W + {\mathcal V}(t),
 \end{eqnarray}
 where the operator, 
 \begin{eqnarray}
  {\mathcal V}(t)=  \sum_{m,n} \lambda_{mn}(t)|m\rangle\langle  n |~,
  \end{eqnarray}
with $m,n = \{0, 1, \dots, 4\}$, describes the influence of the protein-solvent noisy environment.  The noise matrix elements, $\lambda_{mn}(t)$, lead to both relaxation and decoherence processes. In what follows, we restrict ourselves to the diagonal noise effects. (See also \cite{Marcus1,Xu,Ber5}, and references therein.) Then, one can write, 
 \begin{eqnarray}
 \lambda_{mn}(t) = \lambda_n \delta_{mn}\xi(t)~,
 \end{eqnarray}
where $\lambda_n$ is the coupling constant at site, $n$, and $\xi(t)$ is a random process. Generalization to local protein-solvent environments can be done following \cite{NB5}.

 We describe the protein-solvent noisy environment by a random telegraph process (RTP),  $\xi(t) $, with the following properties \cite{Nes1,Nes3},
 \begin{eqnarray}\label{chi_8}
&\langle \xi(t)\rangle =0, \\
 &\langle \xi(t)\xi(t')\rangle = \sigma^2 e^{-2\gamma |t-t'|},
\end{eqnarray}
where $\sigma$ is the amplitude of noise, and $2\gamma$ is the decay rate of the noise correlation function.

The evolution of the  average  components of the density matrix is described by the following system of ordinary differential equations \cite{BNGS,BNLS}:
\begin{eqnarray} \label{IB4}
\frac{d}{dt}{\langle{\rho}}\rangle =i[\langle\rho\rangle,\mathcal H] - \{\mathcal W,\langle\rho\rangle\} - i B\langle\rho^\xi\rangle, \\
\frac{d}{dt}{\langle{\rho^\xi}}\rangle =i[\langle\rho^\xi\rangle,\mathcal H] - \{\mathcal W,\langle\rho^\xi\rangle - i B\langle\rho\rangle- 2\gamma \langle\rho^\xi\rangle , \label{IB5}
\end{eqnarray} with 
\begin{eqnarray}
&& B = \sum_{m,n}(d_m - d_n)|m\rangle \langle n|,\\
&&d_{m}=\lambda_m \sigma,\\
&& \langle\rho^\xi\rangle = \langle\xi\rho \rangle/\sigma.
\label{IB6}
\end{eqnarray}
The average,  $\langle \,\dots \,\rangle$, is taken over the random process. We used the approach developed in \cite{KV1,KV2,KV3} for the RTP to derive Eqs. (\ref{IB4}) and (\ref{IB5}).

Employing Eqs. (\ref{IB4})--(\ref{IB6}), one can show that  the following normalization condition is satisfied,
 \begin{equation}
 	\label{C1}
 	\sum_{n=0}^4 \langle\rho_{nn}(t)\rangle+\eta_0(t)+\eta_4(t)=1.
 \end{equation}
Eq. (\ref{C1})  requires that the total probability of finding the exciton among the five discrete levels and in two sinks is unity for all times. 

\subsection*{Resonant noise} 
The ET rate of the ``donor-acceptor" dimer depends on the noise characteristics of the RTP. In particular, there exist the conditions of ``resonant noise" resulting in a maximal ET rate \cite{NB5,GNB}. According to \cite{NB5}, the ET rate, $\Gamma_{DA}$, is given by:
\begin{equation}
\label{rate}
\Gamma_{DA}={{8\gamma|V_{DA}|^2d^2}\over{(d^2-\varepsilon^2)^2+4\gamma^2\varepsilon^2}},
\end{equation}
where
\begin{equation}
d=(\lambda_D-\lambda_A)\sigma, 
\end{equation}
is the renormalized amplitude of the noise. As one can see from Eq. (\ref{rate}), the rate, $\Gamma_{DA}$, has a maximum,
\begin{equation}
\label{max}
\Gamma_{DA}^{(max)}={{4\gamma|V_{DA}|^2}\over{\sqrt{\varepsilon^4+4\gamma^2\varepsilon^2}-\varepsilon^2}},
\end{equation}
at the ``resonant amplitude" $d_{DA}^{(res)}=(\varepsilon^4+4\gamma^2\epsilon^2)^{1/4}\approx\varepsilon$ (last expression is valid for $2\gamma\ll\varepsilon$, which is assumed in our numerical simulations).

\subsection*{Weakly and strongly coupled chlorophyll dimer}
In the numerical simulations we report on in the next section, we call the chlorophyll-based dimer, $|Chla^*_5\rangle-|Chlb^*_5\rangle\equiv |1\rangle-|2\rangle$ {\em weakly coupled} if $|V_{12}/\delta|\ll 1$, where $\delta=\varepsilon_1-\varepsilon_2$. In this case, two orbitals (eigenfunctions) of the dimer, $|\varphi_-\rangle$ and $|\varphi_+\rangle$, are close to the unperturbed site states, $|\varphi_-\rangle\approx |1\rangle$ and $|\varphi_+\rangle\approx |2\rangle$ \cite{BNGS}. We call the dimer {\em strongly coupled} if $|V_{12}/\delta|\gtrsim 1$.  In this case, the two orbitals of the dimer, $|\varphi_-\rangle$ and $|\varphi_+\rangle$, represent mixtures of the two site states \cite{BNGS}. The corresponding eigenenergies are: $E_{\pm}=(\varepsilon_{1}+\varepsilon_2)/2\pm\sqrt{\delta^2+4|V_{12}|^2 }/2$ $(E_+\geqslant E_-$). Usually, the chlorophyll-based dimer, $|1\rangle-|2\rangle$, is weakly coupled in the CP29 LHC. However, according to \cite{Ahn}, in the NPQ regime this dimer becomes strongly coupled. This transformation of the dimer from weakly to strongly coupled can occur, for example, due to an asymmetric modification (as a result of asymmetric protonation) of local potentials at sites $|1\rangle$ and $|2\rangle$.

\section{Results of numerical simulations}

We analyzed the dependence of NPQ performance on the following parameters: (1) the resonant properties of the protein-solvent environmental noise (controlled by parameter $d$), (2) the coupling strength of chlorophyll-based dimer (controlled by $|V_{12}/\delta|$; we fix $V_{12}$, and vary $\delta$), and (3) the rates to the sinks, controlled by $\Gamma_0$. We numerically computed the solutions of Eqs. (\ref{IB4}) and (\ref{IB5}) for the density matrix components, averaged over noise realizations. In all numerical simulations, we set $\hbar= 1$. Then, the values of parameters in energy units are measured in $\rm ps^{-1}$ ($1\rm ps^{-1} \approx 0.66\rm {meV}$). Time is measured in $\rm ps$.

\begin{figure}[tbh]
	\scalebox{0.25}{\includegraphics{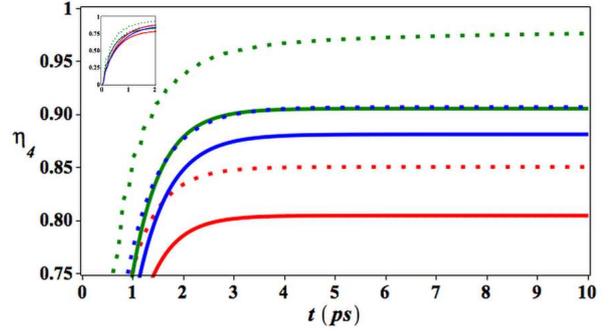}}
	\caption{Time dependence of the NPQ efficiency,  
		$\eta_4(t)$.  Solid/dotted curves demonstrate weakly/strongly coupled chlorophyll dimer. Resonant noise on the transition $|1 \rangle 
		\rightleftarrows |0 \rangle$ ($d_0=-90$): $\Gamma_0=2$ (red), $\Gamma_0=1$ (blue). Non-resonant noise on the transition, $|1 \rangle \rightleftarrows |0 \rangle$ ($d_0=0$): $\Gamma_0=2$ (green). The insert demonstrates the dynamics of the NPQ efficiency for short times. Fixed parameters: $V_{10}=V_{20}=30$,  $V_{12}=15$,  
				$V_{13} = V_{34}=25$, $\varepsilon_0=-90$, $\varepsilon_1=60$,  
				$\varepsilon_1-\varepsilon_2 =\delta $, $\varepsilon_3=45$,  
				$\varepsilon_4=30$, $\Gamma_4 =10$, 
				$\gamma=10$, $d_1= 60$, $d_2 = 90$, $d_3 =45$, $d_4 =30$.  Initial 
				conditions: $|\varphi(0)\rangle=|\varphi_-\rangle$.
		\label{B6a}}
\end{figure}

Noise amplitudes, $d_n$ $(n=0,...,4$), were chosen close to those discussed in the literature. In particular, ``resonant conditions" $(d_n-d_m\approx\varepsilon_n-\varepsilon_m$), discussed in \cite{NB5}, were used for the dimer transitions, $|m\rangle\rightleftarrows |n\rangle$ ($m,n =1,2,3,4$). 
The noise correlation decay rate was $2\gamma=20\, \rm 
ps^{-1}$. The rates to the sinks were: $\Gamma_0=1~{\rm and}~2 \, \rm ps^{-1}$, and 
$\Gamma_4=10 \, \rm ps^{-1}$. 

Fig. \ref{B6a} shows the results of our numerical simulations: the efficiency, $\eta_4(t)$, of the sink $|S_4 \rangle$, which is associated with the beneficial charge-transfer channel. The efficiency, $\eta_4(t)$, saturates at relatively short times, $t_{sat}=(2-4)\,\rm ps$. 
Note, that in the saturation regime, all density matrix elements, 
$\rho_{nm}(t)$  (for discrete states), approach zero, and the total probability 
accumulates in the two sinks. Table \ref{T} provides an overview of the 
resulting efficiency, $\eta_4$. 

\begin{table}[h]
\caption{Efficiency $\eta_4$.}
\begin{tabular}{c}
	\scalebox{0.19}{\includegraphics{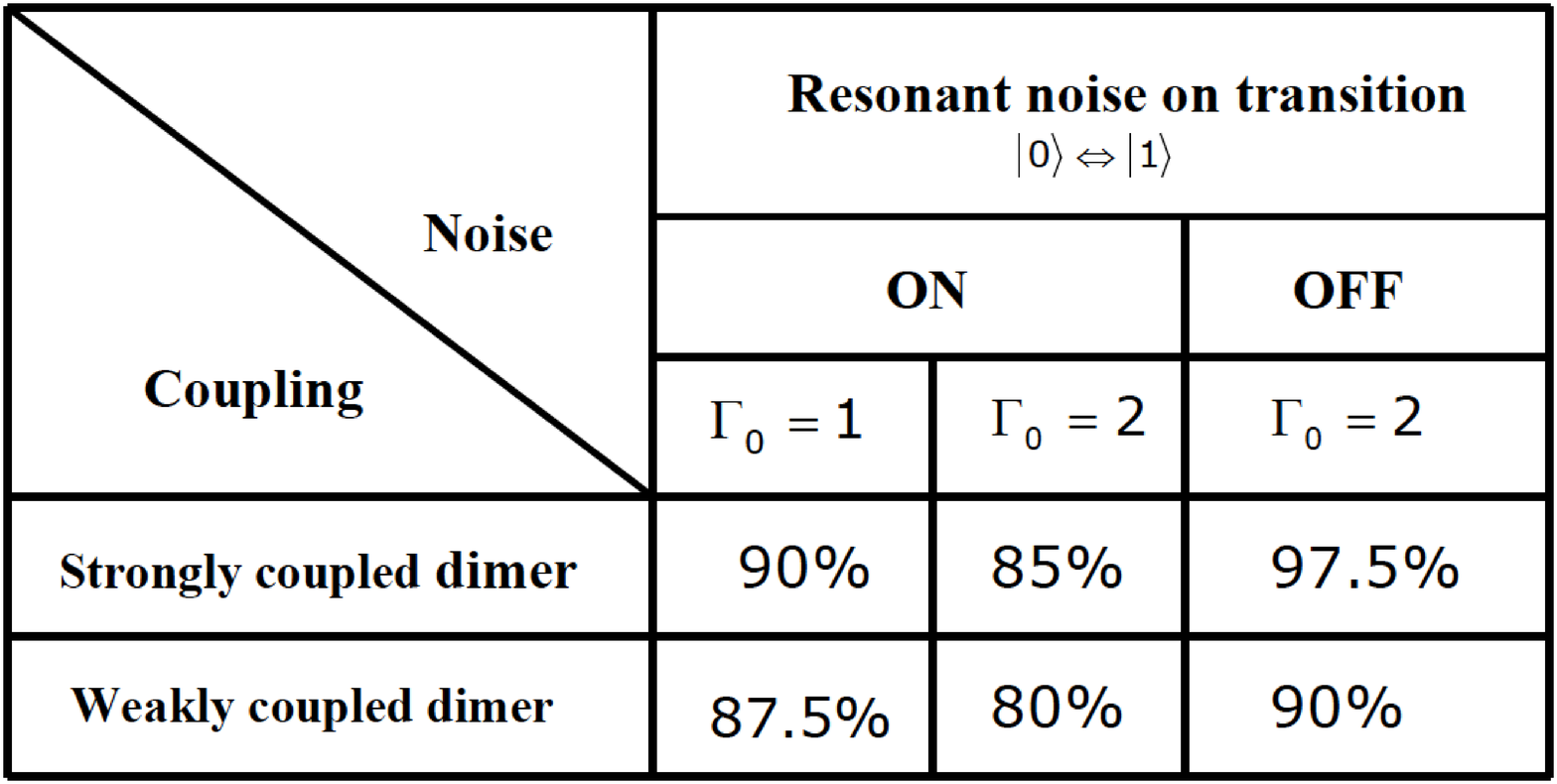}}
\end{tabular}
\label{T}
\end{table}%

\subsection*{Improving the NPQ performance by transition from weakly to strongly coupled chlorophyll dimer} 

The solid red curve in Fig. \ref{B6a} corresponds to the weakly coupled chlorophyll dimer, $|1\rangle-|2\rangle$ ($V_{12}=15$, $\delta=\varepsilon_1-\varepsilon_2=-90$, $|V_{12}/\delta|\approx 0.17$) with a relatively high rate to the damaging sink, $\Gamma_0=2$, and resonant environmental  noise applied to the dimer $|1\rangle-|0\rangle$. In this case, the asymptotic efficiency of the NPQ is relatively small, $\eta_4\approx 80\%$. This means that $20\%$ of the photosynthetic organisms die in this regime. 

Suppose that the weakly coupled chlorophyll dimer, $|1\rangle-|2\rangle$, is transformed in the NPQ regime to be strongly coupled, as assumed in \cite{Ahn}. The result of this transformation is demonstrated  by the red-dotted curve ($V_{12}=15$, $\delta=-15$, $|V_{12}/\delta|=1$). The saturated efficiency approaches, $\eta_4\approx 85\%$. The outcome of the transformation from weakly to strongly coupled dimer thus improves the efficiency, $\eta_4$, by approximately $5\%$. This improvement is the result of destructive interference effects for the probability amplitude of state $|0\rangle$.

Similar effects of improvement of the NPQ efficiency occur when the ET rate, $\Gamma_0$, to the damaging channel is set to $\Gamma_0=1$. Weakly coupled chlorophyll dimer now results in $\eta_4\approx 87.5\%$ efficiency (blue curve in Fig. \ref{B6a}, $\delta=-90$, $\Gamma_0=1$), while strongly coupled chlorophyll dimer is performing $\approx 2.5\%$ better, resulting in $\eta_4\approx 90\%$ (blue-dotted curve, $\delta=-15$, $\Gamma_0=1$). Even though the absolute values of the NPQ efficiencies have increased (in comparison with the case $\Gamma_0=2$), the relative improvement is less.

\subsection*{Improving the NPQ performance by transition from resonant to non-resonant environmental noise}
NPQ efficiency improves by approximately $10\%$ when non-resonant environmental noise is applied to the dimer $|1\rangle-|0\rangle$: weak coupling and $\Gamma_0 = 2$ lead to $\eta_4\approx 80\%$ under resonant noise (red curve), and to $\eta_4\approx 90\%$ under non-resonant noise (green curve). 

The most favorable case of the ones we tried here is strong coupling and $\Gamma_0 = 2$ under non-resonant noise (green-dotted curve), resulting in an efficiency of $\approx 97.5\%$.
We would expect that if we do an optimization over parameters, than the efficiency should be even closer to 100\%.

\subsection*{Restoring NPQ efficiency in response to environmental change}

The results presented in Fig. \ref{B6a} can be used to develop strategies for restoring NPQ efficiency in response to changes in the environment. To summarize, the efficiency, $\eta_4$, is lowest ($\approx 80\%$) with weak coupling, larger $\Gamma_0$, and resonant noise. By switching to strong coupling, and also to lower $\Gamma_0$, the efficiency increases to $\approx 90\%$. There is another way to change the efficiency from 80\% to 90\%, and that is simply by switching from resonant to non-resonant noise. 

This suggests that pH-dependent modifications of the choroiphyll dimer coupling in CP29 could restore NPQ efficiency in response to environment changes. Imagine that the CP29 dimer operates with weak coupling and large $\Gamma_0$, under non-resonant noise, where the NPQ efficiency is $\eta_4\approx 90\%$ (corresponding to the green curve in Fig. \ref{B6a}). Now, suppose that the protein-solvent environment experiences a modification which results in resonant noise on the dimer $|1\rangle-|0\rangle$. This modification increases the ET rate to the damaging channel, $|S_0\rangle$, and, correspondingly, decreases the NPQ efficiency to $\eta_4 \approx 80\%$ (red curve in Fig. \ref{B6a}). This effect could then be counteracted by switching to strong dimer coupling and lower $\Gamma_0$, thereby restoring the efficiency to $90\%$.

As we discussed above, efficiency would be partially restored by a transition from weakly to strongly coupled chlorophyll-based dimer. This transformation increases the NPQ efficiency, $\eta_4$, by $\approx 5\%$ (red-dotted curve, $\eta_4\approx 85\%$). Another way is to keep the weakly coupled dimer, $|1\rangle-|2\rangle$, but to decrease the ET rate, $\Gamma_0$, to the damaging channel, which improves the NPQ efficiency to $\approx 87.5\%$. 

\section {Conclusion}

We demonstrated that the efficiency of the charge-transfer nonphotochemical
quenching in CP29 can be improved by transformation from a weakly to a strongly coupled $Chla^*_5-Chlb^*_5$ dimer. This transformation could be related to the NPQ mechanism by the CTS discussed in \cite{Ahn}. We also demonstrated strategies for restoring the NPQ efficiency when the environment changes. We analyzed numerically only a limited number of the possible system-environment scenarios. Many additional considerations and generalizations can be studied using our approach. The mathematical advantages of our approach are that it (i) allowed us to derive an exact and closed system of ordinary differential equations, which is easy to analyze and to solve numerically and (ii) does not require the use of small parameters and uncontrolled perturbative techniques. We expect that the results discussed in this paper will be useful for a better understanding of the NPQ mechanisms in photosynthetic organisms, and will stimulate new experimental studies. In particular, the experimental verification of possible mechanisms of transformation from a weakly  to a strongly coupled chlorophyll-based dimer, in the NPQ regime, would be of significant interest.  One of the important theoretical problems for future research is to understand how the relationship between (in)efficient use of information and energy dissipation  \cite{Still, Grimsmo, GS2015} comes to bear on the approach presented here.\\ 


This work was carried out under the auspices of the National Nuclear Security Administration of the U.S. Department of Energy at Los Alamos National Laboratory under Contract No. DE-AC52-06NA25396. A.I.N. acknowledges the support from the CONACyT and the suppo345).
rt during his visit of the B-Division at LANL. G.P.B. and R.T.S. acknowledge the support from the LDRD program at LANL. S.S. is grateful for support from the Foundational Questions Institute (Grant No. FQXi-RFP3-1


\begin{thebibliography}{100}
	
\bibitem{Perrine}
Z. Perrine, S. Negi, R.T. Sayre, {\em Optimization of photosynthetic light energy utilization by microalgae}, Algal Research, {\bf 1}, 134  (2012).

	
	\bibitem{book1}
	B. Demming-Adams, G. Garab, and W. Adam III Govindjee (Eds), {\em Non-Photochemical Quenching and Energy Dissipation in Plants, Algae and Cyanobacteria}, (Springer, 2014).
	
	\bibitem{Ahn}
	T.K. Ahn, T.J. Avenson, M. Ballottari, Y.C.  Cheng, K.K. Niyogi,  R. Bassi, and  G.R. Fleming, Architecture of a charge-transfer state regulating light harvesting in a plant antenna protein, Science, {\bf 320}, 794 (2008).	
	
	\bibitem{Dreuw1}
	A. Dreuw,  G.R. Fleming,  and  M. Head-Gordon,  Charge-transfer state as a possible signature of a zeaxanthin-chlorophyll dimer in the non-photochemical quenching process in green plants. J. Phys.
	Chem. B, {\bf 107}, 6500 (2003).
	
	\bibitem{Dreuw2}
	A. Dreuw, G.R. Fleming, and M. Head-Gordon, Role of Electron transfer quenching of chlorophyll fluorescence by carotenoids in non-photochemical quenching of green plants, Biochem. Soc. Trans., {\bf 33}, 858 (2005). 
	
	\bibitem{Ahn2}
	Y.C. Cheng,  T.K.  Ahn,  T.J. Avenson,  D. Zigmantas,  K.K. Niyogi, M. Ballottari, R.  Bassi, and G.R. Fleming,  Kinetic modeling of charge-transfer quenching in the CP29 minor complex, J. Phys.
	Chem. B, {\bf 112}, 13418 (2008).
	
	\bibitem{BNGS}
	G.P. Berman, A.I. Nesterov, S. Gurvitz, and R.T. Sayre, Possible role of interference and sink effects in nonphotochemical quenching in photosynthetic complexes, arXiv:1412.3499 [physics.bio-ph]
	
	\bibitem{BNLS}
	G.P. Berman, A.I. Nesterov, G.V. Lopez, and R.T. Sayre,  Superradiance transition and nonphotochemical quenching in photosynthetic complexes, J. Phys. Chem. C., {\bf 119}, 22289 (2015).
	
	\bibitem{Pudlak1}
	M. Pudlak and R. Pincak, Modeling charge transfer in the photosynthetic reaction center, Phys. Rev. E, {\bf 68}, 061901 (2003).
	
	\bibitem{Pudlak2}
	M. Pudlak, Primary charge separation in the bacterial reaction center: Validity of incoherent sequential model, J. Chem. Phys., {\bf 118}, 1876 (2003).
	
	
	\bibitem{Ber3}
	G. Celardo, F. Borgonovi, M. Merkli, V.I. Tsifrinovich, and G.P. Berman, Superradiance transition in photosynthetic light-harvesting complexes, J. Phys. Chem., {\bf 116}, 22105 (2012).
	
	\bibitem{Ber4}
	D. Ferrari, G.L. Celardo, G.P. Berman, R.T. Sayre, and F. Borgonovi, Quantum biological switch based on superradiance transitions, J. Phys. Chem., {\bf 118}, 20 (2013).
	
	\bibitem{Lloyd1}
	P. Rebentrost, M. Mohseni, I. Kassal, S. Lloyd, and A. Aspuru-Guzik,  Environment-assisted quantum transport, New J. Phys., {\bf 11}, 033003 (2009).
	
	\bibitem{CDCH}
	A.W. Chin, A.~Datta, F.~Caruso, S.F. Huelga, and M.B. Plenio, New J. Phys., Noise-assisted energy transfer in quantum networks and light-harvesting complexes,  \textbf{12}, 065002 (2010).
	

	
	\bibitem{Marcus1}
	R. Marcus and N. Sutin, Electron transfers in chemistry and biology, Biochimica et Biophysica
	Acta, {\bf 811}, 265 (1985).
	
	\bibitem{Xu}
	D. Xu and K. Schulten, Coupling of protein motion to electron transfer in a photosynthetic reaction center: investigating the low temperature behavior in the framework of the spin-boson model, Chemical Physics, {\bf 182}, 97 (1994).
	
	\bibitem{Ber5}
	M. Merkli, G.P. Berman, and S.T. Sayre, Electron transfer reactions: Generalized spin-boson approach, J. Math. Chem., {\bf 51}, 890 (2013).
	
		\bibitem{NB5}
		A .I. Nesterov and G. P. Berman, Role of protein fluctuation correlations in electron transfer in photosynthetic complexes, Phys. Rev. E, {\bf 91}, 042702 (2015).
		
		\bibitem{Nes1}
		A. I. Nesterov, G. P. Berman, and A. R. Bishop, Non-Hermitian approach for modeling of noise-assisted quantum electron transfer in photosynthetic complexes, Fortschritte der Physik, {\bf 61}, 95 (2013).
		
		\bibitem{Nes3}
		A. I. Nesterov, G. P. Berman, J. M. S\'anchez M\'artinez, and R. Sayre, Noise-assisted quantum electron transfer in photosynthetic complexes, J. Math. Chem., {\bf 51}, 1 (2013).
		
	\bibitem{GNB} 
	S. Gurvitz, A.I. Nesterov, and G.P. Berman, Noise-assisted quantum electron transfer in multi-level donor-acceptor system, arXiv:1404.7816 [physics.bio-ph].	

	
	\bibitem{KV1}
	V.~Klyatskin, 	{\em Stochastic Equations through the Eye of the Physicist}, (Elsevier, 2005).
	
	\bibitem{KV2}
	V.~Klyatskin, 	{\em Dynamics of Stochastic Systems}, (Elsevier, 2005).
	
	\bibitem{KV3}
	V.~Klyatskin, 	{\em Lectures on Dynamics of Stochastic Systems}, (Elsevier, 2011).
	
	
	
	\bibitem{Still}
	S. Still, D.A. Sivak, A.J. Bell,  and G.E. Crooks, Thermodynamics of prediction, Phys. Rev. Lett., {\bf 109}, 120604 (2012).
	
	\bibitem{Grimsmo}
	A. L. Grimsmo, Quantum correlations in predictive processes, Phys. Rev. A, {\bf 87} 060302 (2013).
	
	\bibitem{GS2015}
	A. L. Grimsmo and S. Still, Quantum predictive filtering, arXiv:1510.01023 [quant-ph] (2015).
	
	\end{thebibliography}
	\end{document}